\documentstyle[preprint,aps,amsfonts,amssymb,epsfig]{revtex}

\preprint{TH98.2.3-LV6422} \title{Intermediate spin and quantum critical 
points, etc.} \author{S. E. Barnes }

\address{ Department of Physics, University of Miami, Coral Gables, 
Florida 33124 }

\begin{document}
\draft

\date{\today} \maketitle
\begin{abstract}
Unlike that of SO(3) or SU(2), the Lie algebra for SO(2), which 
defines intermediate spin, comprises only $S_{z}$ and implies 
$S^{\pm}$ commute.  In general, $S_{z}$ has a continuous spectrum.  
This intermediate spin scheme can realized for the low energy 
excitations of a wide class of large spin magnets.  A magnetic field 
provides the necessary time reversal symmetry breaking and controls 
the effective value of the spin $\tilde S$.  Physical quantities are 
periodic in the equilibrium magnetization component induced by this 
field.  In particular for one dimensional antiferromagnets there are 
periodic regions on the field axis for which the model is quantum 
critical while in two or three dimensions criticality is reduced to 
points.
\end{abstract}


There is an elementary proof that the usual SO(3) and SU(2) spin 
algebra leads to values of the spin $S$ which are either whole or 
half-integer.  {\it Intermediate spin\/} arises in two dimensions and 
the context of anyons \cite{khare}.  In the absence of time reversal 
invariance, the associated SO(2) algebra admits a continuous spectrum 
for $S_{z}$.  For anyons the spectrum of $S_{z}$ is determined by the 
statistical parameter $\alpha$.  In this Letter it will be shown that 
certain physically realizable large spin magnets with constituent 
whole or half-integer spin $S$, and with a suitably directed 
time-reversal-symmetry-breaking magnetic field, can approximate to the 
intermediate spin scheme with the magnitude of the magnetic field 
controlling the value of the effective statistical parameter $\alpha$.  
The effective spin value $\tilde S = S-\alpha/2 $ is determined by the 
physical value $S$ in zero field and otherwise reflects the spectrum 
of $S_{z} = n- \alpha/2$; $n$ an integer.  Most thermodynamic 
properties are periodic in the field with a period $H_{p}$ determined 
by the Hamiltonian parameters.  Increasing the applied field by 
$H_{p}/2$ will convert a half-integer magnet into its whole integer 
equivalent.  For arbitrary values of $H$ the properties are, in 
general, intermediate between those of half and whole-integer spin.  
It will be shown explicitly for both ferro and antiferromagnetic 
systems with a large easy axis anisotropy energy $D$ that most 
properties are strictly periodic with a period proportional to $D$.  
Antiferromagnets with a small or zero anisotropy energy also exhibit 
intermediate spin behavior for larger fields and sufficiently large 
spin values.  Here the period $H_{p}$ is proportional to the exchange 
energy $J$.  The {\it exceptional\/} thermodynamic quantity, in both 
cases, is the uniform magnetization $M_{0}$.  This is a steadily 
increasing function of $H$ (or $\alpha$) with {\it plateaus\/} 
whenever there is a gaped ``whole integer'' phase.  For a uniform 
system such plateaus are separated by increments in $M_{0}$ which 
correspond to a single $\mu_{B}$ per physical spin.

Intermediate spin systems can have various dimensions.  Zero 
dimensional models corresponding to the tunneling of the magnetization 
of nano-particles.  For {\it zero field\/} Loss et al \cite{loss} and 
von Delft and Henley \cite{henley} have shown the ground doublet 
tunnel splitting is strictly zero for half-integer spin when, for the 
same model and for whole-integer spin, this splitting is finite.  Hard 
axis nano-magnets are explicitly covered by these results and with a 
finite field parallel to the hard axis must also manifest intermediate 
spin properties.  {\it Both\/} on the basis of the general 
intermediate spin considerations{\it and\/} from the direct solution 
of the problem, it is here shown that this tunnel splitting is indeed 
periodic in the field.  Tunneling in antiferromagnets has also been 
studied \cite{seb97,Chiolero}.  It has been demonstrated explicitly 
that the anticipated periodic behavior is exhibited even for systems 
with a small or zero explicit anisotropy energy.

For one dimension and {\it zero field}, Haldane \cite{hal} has 
conjectured, for antiferromagnetic chains for which the whole integer 
spin version exhibits a gap, the half-integer equivalent does not.  
Hard axis such spin chains should conform with this conjecture but in 
a field again must also reflect intermediate spin properties.  Using 
well established results, the large anisotropy limit can be solved.  
On this basis it can again be {\it explicitly\/} demonstrated that a 
certain field $H_{p}/2$ converts a whole into half-integer spin 
problem.  Whatever the whole or half-integer nature of the physical 
spins, these models have gapless ``half-integer'' quantum critical 
regions surrounding periodic points on the field axis of their phase 
diagram.  The rest of the field axis comprises gaped ``whole integer 
phases''.  Results for antiferromagnets chains with a small or zero 
anisotropy energy are more difficult to demonstrate.  However the work 
of Oshikawa \cite{Oshikawa} and Totsuka \cite{Totsuka} on smallish 
half-integer spin chains shows that ``whole integer'' plateaus exist 
thereby supporting the general idea to be present here that a magnetic 
field leads to spin transmutation.

{\it In two or three dimensions}, where the Haldane conjecture does 
not apply, there are similar periodic phases diagrams with, e.g., 
half-integer phases which terminate with field determined quantum 
critical points.  Given the current high level of interest in quantum 
critical points the ready availability of high fields this suggests any 
number of exciting experimental possibilities.

That intermediate spin properties can arise in physical whole or 
half-integer spin systems has implications for certain concepts based 
on topological considerations.  Both the results of Loss et al 
\cite{loss} and von Delft and Henley \cite{henley} and the Haldane 
conjecture \cite{hal} are explained \cite{frad} in terms of the so 
called {\it topological term\/} in the effective action.  In the usual 
notation this is $iS(1-\cos \theta)\dot \phi$ and the differences 
between whole and half-integer spin flow from the value of the spin 
$S$ in this expression.  These results and conjectures correspond to
Hamiltonians with time reversal symmetry while the realization of 
intermediate spin implies this symmetry is broken.  It must be that 
the same interference effect which destroy the gaps for half-integer 
spin {\it can and do\/} occur periodically as a function of applied 
field even for systems constituted of physical whole-integer spins.

The necessary dimensional reduction SO(3) or SU(2) $\Rightarrow$ SO(2) 
is most easily accomplished by a hard axis anisotropy which forces 
spins to lie in the $x-y$-plane.  However for {\it antiferromagnets\/} 
the field alone suffices.  As is well known, for ordered phases, a 
magnetic field $H_{sf}$ causes a ``spin-flop'' transition.  What is 
perhaps not appreciated is that this can be considered as a 
competition between the true anisotropy energy and an effective hard 
axis anisotropy energy $D_{H} \sim H^{2}/J$, $J$ the exchange, induced 
by the external field $H$.  With $\vec H$ along the easy axis the 
spin-flop occurs when $D_{H} \sim D$ a criterion which correctly leads 
to $H_{sf} \sim \sqrt{DJ}$.  For the intermediate spin scheme to apply 
it must be that this induced hard axis anisotropy $D_{H}$ is larger 
than the exchange, i.e., that $H > J$.  When this inequality is well 
obeyed, and for large enough spin, most thermodynamic properties are 
again periodic with plateaus in the magnetization.

It is again emphasized that the models which approximate to the 
intermediate spin scheme are constituted of whole and/or half-integer 
spins $S$ and that the point of the exercise is no more to disprove 
the basic theorem on spin values than the demonstration that condensed 
matter systems can exhibit the properties of anyons is intended to 
disprove that these systems are ultimately composed of bosons and 
fermions.  The Lie algebra of SO(2) has only the single operator 
$S_{z}$ which is the generator of rotations about the $z$-axis.  
Consider a rotational degree of freedom $\phi$ of two anyons 
\cite{khare}.  In a non-singular gauge define: $S_{z} ={1\over 2}\left( - 
i(d/d\phi ) - \alpha\right)$, where $\alpha$ is the statistical 
parameter and where the eigenfunctions are $F_{\ell} = e^{2is\phi}, \ s 
= 0, \pm 1, \pm 2 \ldots$ to give a spectrum $S_{z} = s - \alpha/2$.  
The raising and lowering operators are simply $S^{\pm} = \tilde S 
e^{2i\theta}$, where $\tilde S$ is spin value to be defined below.  The 
spin algebra is then uniquely:
\begin{equation}
[S_{z}, S^{\pm} ] = \pm S^{\pm} 
\ \ \ \
{\rm and}
\ \ \ \
[S^{+},S^{-} ] = 0
\label{uno}
\end{equation}
These commutation rules, which are equivalent to $[\phi,p_{\phi}] = 
i\hbar$ and $[\phi,\phi] = 0$, are here taken as the {\it 
definition\/} of intermediate spin.  Time reversal symmetry, ($S_{z} 
\to - S_{z}$) implies that the spectrum of $S_{z}$ is whole or half 
integer, however in general this is {\it not\/} the case, i.e., this 
spectrum is continuous.  An intermediate spin model comprises a 
Hamiltonian $\cal H$ given as a function of $S_{z}$ and $S^{\pm}$.  It 
is then observed that if $-\alpha/2$ is the smallest absolute value in 
the applicable spectrum of $S_{z}$, this Hamiltonian will only couple 
eigenstates with $S_{z}= n - \alpha/2$ where $n$ is an integer.  
Finally, the spin value $\tilde S$ is {\it chosen\/} to be $S-\alpha$, 
where $S$ is an integer.  (A different choice might be absorbed into a 
re-definition of the interaction parameters in the transverse part of 
$\cal H$>) This $\tilde S$ is an eigenvalue of $S_{z}$ and 
characterizes a solution of $\cal H$.  With physical spin $S$ models 
$\tilde S = S$ when $\alpha = 0$.

Given that Eqns.  (\ref{uno}) define intermediate spin, it is 
necessary to show that this represents an interesting possibility in 
two ways.  First that there real physical systems composed out of 
whole and/or half-integer spins for which the intermediate spin 
description is valid as a good approximation, and second that such 
systems have the non-trivial properties described above.

That this is the case is easiest to demonstrate for the above 
mentioned class of {\it hard axis\/} magnets with not {\it too\/} 
small values of the constituent whole or half-integer spin $S$ 
\cite{size}.  The role of the large anisotropy energy is to enforce 
the dimensional reduction discussed above.  The hard direction is 
taken to be the $z$-axis and the class of Hamiltonians is defined by:
\begin{equation}
{\cal H} =
g \mu_{B} H \sum_{n} S_{z,n}
+
D \sum_{n} {S_{z,n}}^{2}
+
{\cal H}_{1}(\vec S_{n})
\label{un}
\end{equation}
where $D$ is a {\it positive\/} anisotropy energy which is 
sufficiently large that
\begin{equation}
S^{2} D \gg\, g \mu_{B} H	,
	\label{deuxa}
\end{equation}
and
\begin{equation}
S^{2} D\gg\, <{\cal H}_{1}>,
	\label{deuxb}
\end{equation}
and where the interaction Hamiltonian ${\cal H}_{1}$ is a fairly 
general scalar function of the $\vec S_{n}$, see below.  The 
expectation value is for the low lying states of interest and the 
indice $n$ labels the constituent spins each of magnitude $S$.  It is 
trivial that the magnetic field can be absorbed into the anisotropy 
term via a redefinition of the spin operator
\begin{equation}
\hat S_{n,z} \Rightarrow \hat  S_{n,z} - {\alpha \over 2}\, ; \ \ \ \ \
\alpha = 
{g \mu_{B} H \over  D}
\label{quatre}
\end{equation} 
and an unimportant shift in energy.  The spins are forced to lie near 
the $x-y$-plane and, provided the {\it physical\/} spin value is 
sufficiently large, i.e., if
\begin{equation}
S \gg \alpha,
\label{cinq}
\end{equation}  
for low energy states, the matrix elements of $S^{\pm}_{n}$ can be 
replaced by $S$, which is equivalent to the commutation rule 
$[S^{+}_{n},S^{-}_{n^{\prime}}] = 0$ of Eqn.  (\ref{uno}).  Clearly 
the substitution $\hat S_{n,z} \Rightarrow \hat S_{n,z} - \alpha/2$ 
has no effect on $[S^{+}_{n},S^{-}_{n^{\prime}}] = 0$ and leaves 
unchanged $[\hat S_{n,z},S^{\pm}_{n}] = \pm S^{\pm}_{n}$.  If the 
original basis is re-labeled $|S_{n,z}>\Rightarrow |S_{n,z}-\alpha/2>$ 
then $ \hat S_{n,z} |S_{n,z}-\alpha/2> = (S_{n,z} - \alpha/2) 
|S_{n,z}-\alpha/2>$ and the whole algebra is that for intermediate 
spin with an effective spin
\begin{equation}
\tilde S = S - \alpha/2.
\label{sept}
\end{equation} 

There are necessarily some restrictions on the interaction term ${\cal 
H}_{1}$.  Uniform quadratic terms such as $\sum_{n, n^{\prime}} 
J^{\parallel}_{n, n^{\prime}} S_{n,z}S_{n^{\prime},z} \Rightarrow 
\sum_{n, n^{\prime}}J^{\parallel}_{n, n^{\prime}} 
S_{n,z}S_{n^{\prime},z} + h_{1} \sum_{n} S_{n,z}+ {\rm const.}$, as 
$\hat S_{n,z} \Rightarrow \hat S_{n,z} - \alpha/2$ and cause no 
problems since the $h_{1} = z J^{\parallel}/2 \ll g\mu_{B}H$, $z$ the 
number of neighbors, results only in a small modification to the 
definition of the effective applied field.  The corresponding 
transverse terms do not change, i.e., (1/2) $\sum_{n, n^{\prime}} 
J^{\perp}_{n, n^{\prime}} (S_{n}^{+}S_{n^{\prime}}^{-} + H.c.)$ 
transforms to itself.  Similarly for an isotropic expression $\sum_{n, 
n^{\prime}} J_{n, n^{\prime}} (\vec S_{n} \cdot \vec 
S_{n^{\prime}})^{n}$ there are unimportant small modifications to both 
the field and the anisotropy energy.  However, e.g., 
$S_{n,z}(S^{+}_{n^{\prime}}+S^{-}_{n^{\prime}})$, which reflects a 
lowering of symmetry, generates an apparent transverse applied field 
and implies that a small component of the applied field must be used 
to cancel such terms.  However not all terms which lower the symmetry 
are problematic.  In particular the expression $K_{\perp}\sum_{n} 
S_{n,x}^{2}$, which is important for tunneling problems transforms 
into itself.  Thus, while perhaps not a completely general function of 
the spin operators, the allowed operators ${\cal H}_{1}$ cover a wide 
variety of interesting problems.

Turning to the more general case when the anisotropy energy is small 
or zero, it is observed that $\hat S_{n,z} \Rightarrow \hat S_{n,z} - 
\alpha/2$ can apparently still be used to eliminate the magnetic field 
terms since as effectively noted above $J \sum \vec S_{n} \cdot \vec 
S_{n^{\prime}} \Rightarrow J \sum \vec S_{n} \cdot \vec 
S_{n^{\prime}}+ h_{1} \sum_{n} S_{n,z}+ {\rm const.}$ {\it 
Unfortunately\/} in the absence of the large anisotropy term there is 
no assurance that the matrix elements of $S^{\pm}$ reduce to $S$ as 
was the case above.  An appropriate, zero dimensional, tunneling 
problem has been studied in some detail \cite{seb97,Chiolero} and the 
same principles can be applied to arbitrary dimensions.  An abbreviated 
fashion to obtain the their results goes as follows: Effectively the 
model comprises two spin interacting via a $J \vec S_{1} \cdot \vec 
S_{2}$ so the ground state energy is $F(m) = (J/2)[m(m+1) - 2 S(S+1)] 
- m (g \mu_{B} H)]$ where $m$ is the $z$-quantum number for the total 
magnetisation.  The aim of the intermediate spin transformation $\hat 
S_{n,z} \Rightarrow \hat S_{n,z} - \alpha/2$ is to eliminate the 
time-reversal-breaking terms and as a result the equilibrium 
magnetization $g \mu_{B} m_{0}$ is automatically $ g \mu_{B} 
\alpha/2$, i.e., the statistical parameter is determined directly by 
the equilibrium magnetization.  For the tunneling problem the 
equilibrium magnetization gives $\alpha/2 =(g \mu_{B} H/J) - (1/2)$, 
and a ground state energy which contains a $\sim - [(g \mu_{B} 
H)^{2}/2J]$ field generated anisotropy energy.  That there is a 
$\pi/2$ phase shift corresponding to the $1/2$ is confirmed by both 
direct calculation and numerical solution \cite{seb97,Chiolero}.  
These results also show that the true intermediate spin regime, with 
this phase shift, only sets in for larger fields such that $g \mu_{B} 
H$ is a few times $J$ illustrating the necessary development of the 
field induced hard axis anisotropy.  The result $H_{p} = (J / g 
\mu_{B})$ also agrees with direct calculation.  Rather generally, 
antiferrromagnet systems in a finite field have a larger transverse 
that longitudinal susceptibility, develop a field induced hard axis 
anisotropy energy, and undergo spin-flop transitions.  When this 
energy is large compared with the exchange energy the system will 
exhibit intermediate spin properties with in general
\begin{eqnarray}
\alpha = 2 m_{0} = 2{ \chi H \over g \mu_{B}}
\label{cent}
\end{eqnarray}
where $\chi$ is the magnetic susceptibility. While in general, as a 
function of $H$, e.g., whole integer plateaus will be phase shifted as 
compared with the $H=0$ point, such plateaus and half-integer points 
will always be exactly displaced by units of $\mu_{B}/2$ along the 
$m_{0}$ axis. For low fields some plateaus might be missing because a 
large enough anisotropy energy has not yet been developed and the 
systems is not in the intermediate spin regime.

The basic result of this Letter is the observation that, with some 
qualifications on the interaction and when the intermediate spin 
inequalities are well satisfied, many large spin magnets have physical 
properties which are periodic in the equilibrium magnetization.  The 
effective value of the spin $\tilde S = S-\alpha/2$ where the 
statistical parameter $\alpha$ is determined by the equilibrium 
magnetization via the applied field.  Increasing the equilibrium 
magnetization by $\mu_{B}$ per spin maps the problem back to itself 
while an increase by $\mu_{B}/2$, e.g., converts a half-integer 
problem into its whole integer equivalent

As an example, consider first the zero dimensional problem of a small 
easy plane ferromagnetic nano-particle.  This is modeled by a single 
large spin subject to the external and anisotropy fields\cite{chud}, 
i.e., the Hamiltonian is ${\cal H} = g \mu_{B} H S_{z} + K_{\parallel} 
{S_{z}}^{2} + K_{\perp}{S_{x}}^{2}$ where without loss of generality 
it is assumed that $|K_{\parallel}|>|K_{\perp}|$.  (Here $K_{\parallel} \equiv 
D$.)  For an easy plane magnet $K_{\parallel} > 0$.  This 
is of the class of Eqn.  (\ref{un}) with a single physical spin $S$ 
and ${\cal H}_{1} = K_{\perp}{S_{x}}^{2}$.

The problem is formulated \cite{seb} in terms of (Abrikosov) auxiliary 
particles.  A basis $| S_z> \equiv |n>$ is chosen.  Then an auxiliary 
particle, a fermion $f^{}_n$, is associated with each state via the 
mapping $|n> \to f^\dagger_n |>$ where $|>$ is a non-physical vacuum 
without any auxiliary particles.  Defined is a bi-quadratic version of 
an operator $\hat O$ via: $ \hat O \to \sum_{n,n^\prime} f^\dagger_n 
<n|\hat O|n^\prime> f^{}_{n^\prime}$.  The constraint $ Q = \sum_n 
\hat n_n = \sum_n f^\dagger_n f^{}_n =1 $.  It has been shown 
\cite{seb} that such schemes preserve all operator multiplication 
rules including commutation rules.  The replacement rule is applied to 
the Hamiltonian ${\cal H}$ to yield, setting $ h = g \mu_{B} H $:
\begin{eqnarray}
&{\cal H} = \sum_n \Big( K_{\parallel} {n}^2- n h + \nonumber\\
&{1\over 4} 
K_{\perp}[ \left( M_{n}^{n+1}\right)^{2} + 
\left(M_{n}^{n-1}\right)^{2}] \Big) f^\dagger_n f^{}_n\nonumber\\
&+ 
{1\over 4} K_{\perp} \sum_n M_{n}^{n+1} M_{n+1}^{n+2} ( 
f^\dagger_{n+2} f^{}_n + H.c.  ){,}
\label{neuf}
\end{eqnarray}
where the $M_{n}^{n+1} =[S(S+1) - n(n+1)]^{1/2}$ are the matrix 
elements of $S^{\pm}$.  This is {\it two\/} disconnected tight binding 
chains of spinless fermions $f^\dagger_n$.  The constraint $Q=1$ 
implies this is a single particle problem.  The ``site'' indices are 
whole or half-integers following the nature of the constituent spins.

The intermediate spin regime requires inequality (\ref{deuxb}) which 
translates to $K_{\parallel} > K_{\perp}$.  However usually considered 
by others \cite{loss,henley,garg} is the ``tunneling regime'' which 
implies $ K_{\parallel} < S^{2} K_{\perp}$ so large values of $S$ are 
implied.  The wave-function is confined near $n=0$ and the problem is, 
to an excellent approximation, that of a single particle in a harmonic 
well.  The spin value is reflected only by the matrix elements 
$M_{n}^{n+1} \approx S$, again to a very good approximation.  It is 
interesting that there is ``no under the barrier'' tunneling in this 
representation of the problem.  The small ``tunneling'' energy 
difference displayed below is actually the difference in energy 
between the ground states on the two chains.  It arises because the 
harmonic problem on a discrete lattice depends at an exponential level 
on the disposition of the sites relative to the origin of the harmonic 
potential.  In general this disposition is different for these two 
chains.

It is easily appreciated that (i) the only difference between the 
whole and half-integer problems is a shift in the ``sites'' by $1/2$ 
and (ii) an exactly similar shift $h/2K_{\parallel}$ is induced (with 
and unimportant shift in the total energy) by an applied field.  It 
follows that the applied field causes the effective spin value to be 
continuous and physical properties to be periodic as described above.  
This problem has been solved directly using the above described 
auxiliary particle formulation \cite{seb97}.  The {\it result\/} for 
the tunnel splitting is:
\begin{eqnarray}
\delta E &=& 4 |\cos ( \pi \tilde S)|
\sqrt{2 \over \pi} \omega_{0}S_{f}^{1/2}e^{-S_{f}};\nonumber\\
 S_{f} &=&  
2 S\sqrt{(K_{\perp}/ K_{\parallel})}; \ \ 
\omega_{0} = S\sqrt{K_{\perp}/ K_{\parallel}},
\label{onze}
\end{eqnarray}
where $\tilde S = S-\alpha/2$.  This is {\it identical} to the zero field 
results \cite{loss,henley} {\it except\/} for the anticipated change 
$S \to \tilde  S$.  (Garg \cite{garg} had previously shown the tunnel 
splitting is quasi-periodic in the field.  However his development and 
interpretation of this fact was quite different and he did not give an 
analytic result for $\delta E$ which might be compared with the zero 
field result.)

As already mentioned, the tunneling problem for an antiferromagnet 
with a {\it small or zero\/} anisotropy energy has also been solved 
and exhibits the expected oscillations with period $H_{p} = J/g 
\mu_{B}$ \cite{seb97,Chiolero}, in the present notation.

For one dimensional chains and a large easy axis anisotropy, a simple 
case corresponds to the general Hamiltonian Eqn.  (\ref{un}) with 
${\cal H}_{1} = J \sum_{n} \vec S_{n} \cdot \vec S_{n+1}$; $J>0$, 
i.e., an antiferromagnetic Heisenberg model {\it but\/} with a large 
anisotropy energy.  Here the inequality (\ref{deuxb}) is $D > J$.  
Unfortunately there is no available solution unless $D \gg S^{2} J$, 
corresponding to a limiting or fixed point.  In this limit well 
established solutions can can be used to infer the relevant phase 
diagram both for the vicinity of the fixed point and less directly for 
the more general case \cite{gap} when $K_{\parallel} > J$.  A similar 
fixed point approach has been shown useful for the tunneling problem 
\cite{seb97,sebetal}.

It is necessary to add a {\it real space site\/} index $n$ to the 
Fermi operators: $f^{}_{n,m}$.  In the limiting case, $D \gg S^{2} J$, 
zero field, and {\it whole integer spin\/} the ground state of the 
{\it dominant\/} anisotropy term, in zero field for each real space 
site, is $|S_{nz} = 0> = f^{\dagger n}_{0}|>$.  {\it The interaction 
is a perturbation\/} and {\it in the limit\/} the ground state is 
simply $\prod_{n} f^{\dagger n}_{0}|>$, reflecting a singlet at each 
site.  There is a, trivial and large, ``Haldane gap'' in the 
excitation spectrum of magnitude $\sim K_{\parallel}$ \cite{gap}.

In contrast {\it for half-integer spin\/} there is a doublet ground 
state of $K_{\parallel} {S_{nz}}^{2}$ comprising $|S_{nz} = \pm 1/2> = 
f^{n\dagger }_{\pm 1/2}|>$ for each site and for $K_{\parallel} \gg 
S^{2} J$ and small fields, $\alpha < 1$, the effective Hamiltonian 
for these doublets is
\begin{eqnarray}
 &{\cal H}_{a}
=
 g \mu_{B} H \sum_{n}s_{n,z}  \nonumber\\
+
& J \sum_{n} \left[
s_{n,z}s_{n+1,z} + S (s_{n}^{+}s_{n+1}^{-} + H.c.)
\right]
\label{douze}
\end{eqnarray}
where $s_{n,z} = (1/2) [f^{\dagger }_{n, 1/2}f^{}_{n, 1/2} - 
f^{\dagger }_{n,-1/2}f^{}_{ n,-1/2}]$ and $s_{n}^{+} = f^{\dagger 
}_{n, 1/2}f^{}_{ n,-1/2}$.  These operators correspond to a regular 
SU(2) spin algebra with $S=1/2$ and ${\cal H}_{a}$ is the integrable 
anisotropic Heisenberg model {\it without\/} an explicit anisotropy 
energy {\it but\/} with $J_{\perp} / J_{\parallel} =S$.  For small 
fields $H$ this model is gapless and {\it critical\/} \cite{bog}.  All 
correlation functions decay as simple power laws \cite{bog}, i.e.,
\begin{eqnarray}
<s_{n+1,z}s_{n,z}> - <s_{z}>^{2} &\sim& {1 \over n^{\theta}}; \nonumber\\ 
<s_{n+1}^{+}s_{n}^{-}> &\sim& {1 \over n^{1/\theta}}
\label{treize}
\end{eqnarray}
where for this small field regime $\theta = (\pi /2\eta)(1+ \alpha_{1} 
H^{2})$; $\alpha_{1} =(1-(2\eta/\pi))^{2}[\cot(\pi\eta/(\pi - 2 
\eta))/8 \eta \sin^{2}2\eta]$ and where the anisotropy parameter 
$\eta$ is defined by $\cos 2\eta = J_{\parallel}/J_{\perp}= (1/S)$ and 
is of the order of, but slightly greater, than $\pi/4$.

There is \cite{bog} a {\it critical field\/} defined by $H_{c} = J 
\sin^{2} \eta$ and for fields less that $H_{c}$ the system remains 
critical.  Approaching $H_{c}$, $\theta = 2 + 4(\pi \tan \eta \tan 
2\eta)^{-1} \sqrt{H_{c} - H}$ while for $H>H_{c}$ the magnet is 
ordered ``ferromagnetically''.  

Turning to larger finite fields, the model can be equally well be 
solved in the vicinity of fields which are multiples of 
$(H_{p}/2)=K_{\parallel}/g\mu_{B}$.  Specifically for half-integer 
spin when $H=K_{\parallel}/g\mu_{B}$, at each site the state 
$f^{\dagger }_{n,1/2}|>$ is the ground state of the anisotropy energy 
with $f^{\dagger }_{n,-1/2}|>$ and $f^{\dagger }_{n,3/2}|>$ 
degenerate, at an energy $K_{\parallel}$ higher.  For low energies the 
problem maps exactly onto the whole integer equivalent with again the 
large gap $\sim K_{\parallel}$.  There is, of course, a ferromagnet 
moment of $\sim g \mu_{B} /2$ per site.

Doubling the field so $H=2K_{\parallel}/g\mu_{B}$, causes the states 
$f^{\dagger}_{ n,1/2}|>$ and $f^{\dagger }_{n,3/2}|>$ to become 
degenerate so that again to a good approximation the Hamiltonian 
(\ref{douze}) describes the situation except $H \to H - 
2K_{\parallel}/g\mu_{B}$ and $s_{n,z} = (1/2) [f^{\dagger }_{n, 
3/2}f^{ }_{n, 3/2} - f^{\dagger }_{n,1/2}f^{ }_{n,1/2}]$ with 
$s_{n}^{+} = f^{\dagger }_{n, 3/2}f^{}_{ n,1/2}$.  The model is 
critical in the interval $2K_{\parallel}/g\mu_{B} - H_{c}$ to 
$2K_{\parallel}/g\mu_{B} + H_{c}$ and the only significant difference 
with zero field is the existence of an average ferromagnetic moment of 
$\sim g\mu_{B}$ per site.

With each level crossing at $H=2nK_{\parallel}/g\mu_{B}$ the 
model maps back to that in zero field except for a moment of 
$ng\mu_{B}$ per site, while for fields of 
$H=(2n+1)K_{\parallel}/g\mu_{B}$ the map is to whole integer spin with 
an additional moment of $\sim (2n+1)g\mu_{B}/2$.  There is a 
``half-integer'' critical phase delimited by the points 
$2nK_{\parallel}/g\mu_{B} \mp H_{c}$.  This is illustrated in Fig.  1.  
The phase diagram for constituent whole integer spin is similar with a 
shift of $K_{\parallel}/g\mu_{B}$ along the field axis and with 
background moments $\sim (2n+1)g\mu_{B}/2$ around the effective 
half-integer points.

The solutions near the half-integer points connect smoothly with the 
singlet ground states valid near the center of the whole-integer 
phases.  For example, still for a half-integer chain, for fields 
modestly larger than $H_{c}$ the ``ferromagnetic'' ground state from 
the exact solution is, in the current language, approximately 
$\prod_{n} f^{\dagger }_{n,1/2}|>$ which is just the singlet ground 
state for the next whole-integer phase, providing a connection 
between the two solutions.  It is less obvious that Fig.  1 also 
describes the phase diagram when only the weaker inequality 
$K_{\parallel} > J$ is satisfied.  The periodic behavior originates 
from level crossing at fields which are a multiples of $H_{p}$.  The 
inequality guarantees that the ground and low lying states 
$\psi_{\alpha}$ are admixtures of states $ |n,m>$ with small 
coefficients for states with $|m| \sim S$ because these cost too much 
energy.  The change in these $\psi_{\alpha}$ when $H \to H + H_{p}$ 
can be accounted for to a good approximation by the change $|n,m> \to 
|n,m-1>$.  This way of re-stating that a finite field is equivalent to 
a translation in the origin of the potential strongly suggests that 
the weaker inequality is sufficient for the models to exhibit periodic 
behavior, providing one stays within the ``large-D'' phase \cite{gap}.

In two or three dimensions there are rather too many possibilities to 
be enumerated here, however the basic structure along the $H$ axis of 
the phase diagram must always be very similar.  If, e.g., the system 
is composed of half-integer spins then further ``half-integer phases'' can 
occur whenever the magnetic field increases the uniform magntization 
by $\mu_{B}$ per spin.  Well within the intermediate spin regime these 
points will be periodic in the field.  Midway between these 
half-integer phases are ``whole-integer phases''.  There remains only 
critical {\it points\/} at the limits of the ``half-integer phases''.  

The equivalence between the zero field $S=1/2$ half-integer problem 
and a whole integer spin model with an large anisotropy energy and a 
suitable magnet field is not new.  For spin $S=1$ this possibility was 
analyzed at the molecular field level long ago \cite{Tsuneto,Tachiki}.  
Refs.  \cite{Oshikawa,Totsuka} on the other hand show some small spin 
half-integer spin systems exhibit magnetization plateaus. Both sets of 
results can be interpreted as supporting the idea of the transmutation 
of spin described here.

Experimental examples of the predicated periodic behavior are perhaps 
difficult to find because of the required large fields and spin 
values.  Current laboratory fields of $H_{\rm max} \sim 30$T implies 
low temperature magnets.  Small values of the exchange are not 
particularly sought after but certainly exist in the region $100$mK to 
10K making the search for intermediate spin behavior possible with 
current facilities.  The evident experimental signature of 
intermediate spin is the periodic behavior of nearly all thermodynamic 
quantities and in particular the field derivative of the magnetization 
and the specific heat.  The only example of such a quasi-periodic 
behavior of which the author is aware is that of Taft et al 
\cite{taft} for Fe${}_{10}$ which is a zero dimensional example with 
small anisotropy.  For Mn${}_{12}$, Thomas et al., and Friedman et al, 
\cite{thomas} observe periodic behavior but for quite different 
reasons.  Magnetization plateaus which might be interpreted as whole 
$\Leftrightarrow$ half-integer mutations have been observed for small spin 
systems: see \cite{Oshikawa,Totsuka,Tsuneto,Tachiki}.


%

\begin{figure}
\caption{ The zero temperature field axis of the phase diagram for the 
half-integer constituent spins.  The ``half-integer'' phases are 
centered at $H=0,\, (2D/g \mu_{B}),\, (4D/g \mu_{B})\ldots$ while the 
``whole-integer'' phases have their centers at $H= (D/g \mu_{B}),\, 
(3D/g \mu_{B})\ldots$.  The former phases have the relatively small 
width $2H_{c}$ while latter occupy the rest of the phase diagram.  The 
magnetic moments at the center of the phases are $0, \, g\mu_{B} \, 
2g\mu_{B} \ldots$ and $ g\mu_{B}/2 \, 3g\mu_{B}/2 \ldots$ 
respectively.  However the field derivative $\partial M/\partial H$ 
and most other thermodynamic quantities are periodic along this axis.}
\label{C}
\end{figure}

\vfill\eject
\epsfig{width=0.95 \linewidth,file=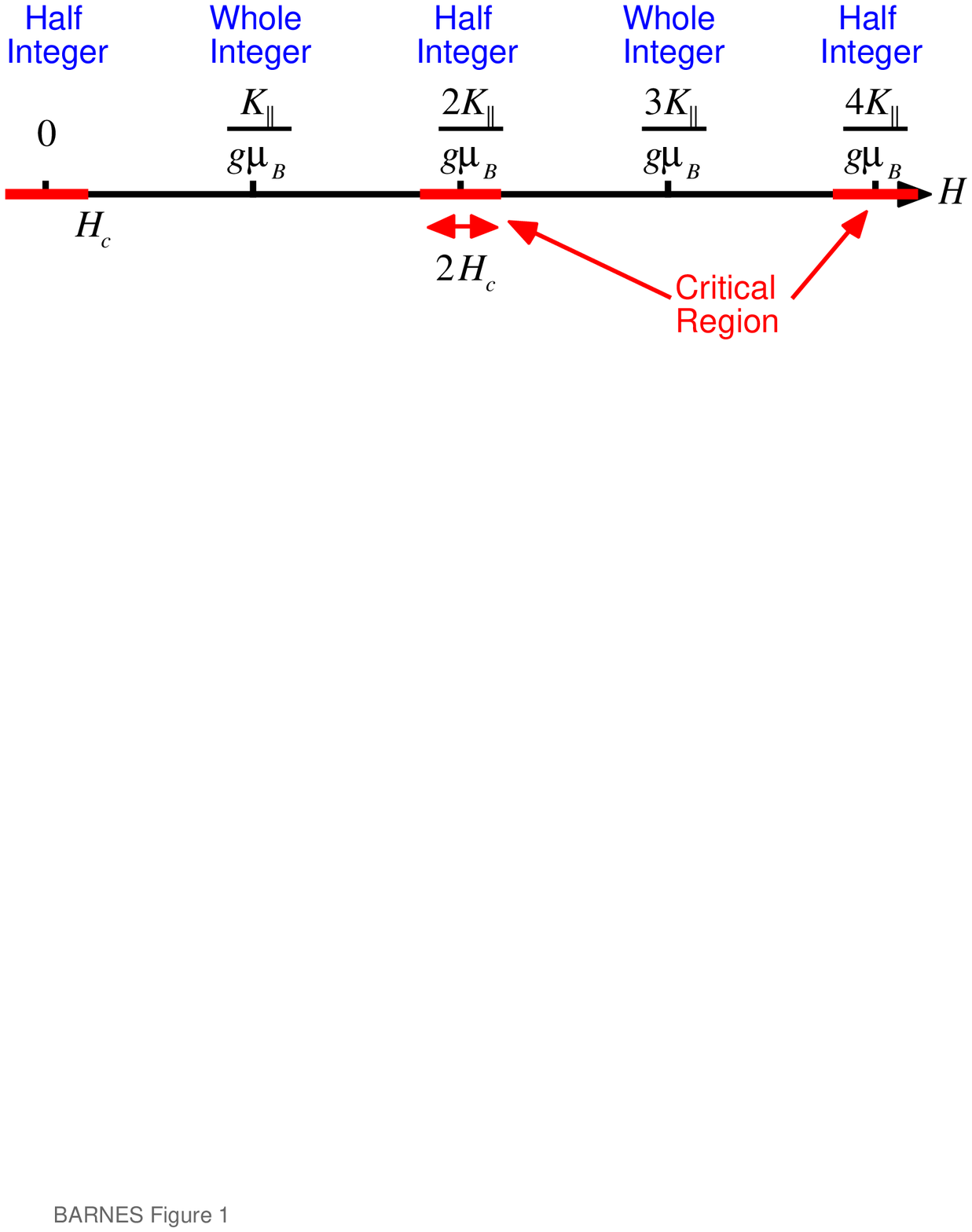}


\begin{references}

\bibitem{khare} See e.g., A. Khare {\it Fractional statistics and 
quantum field theory}, World Scientific, Singapore (1997).


\bibitem{loss} 
D. Loss, D. P. DiVincenzo and G. Grinstein, Phys. Rev. Lett.  
{\bf 69}, 3232 (1992).

\bibitem{henley} 
J. von Delft and C. L. Henley, Phys.  Rev.  Lett.  {\bf 69}, 3236 
(1992).

\bibitem{hal} 
F. D. M. Haldane, J. Appl. Phys. {\bf 57}, 3359 (1985).

\bibitem{Oshikawa}
M. Oshikawa, M. Yamanaka, and I. Affleck, Phys. Rev. Lett. {\bf 78},
1984 (1997).

\bibitem{Totsuka} K. Totsuka, Phy. Rev. B{\bf 57}, 3454 (1998).

\bibitem{seb97} S. E. Barnes, cond-mat/ 9710302, (1997).

\bibitem{Chiolero} A. Chiolero and D. Loss, Phys. Rev. Lett. {\bf 80}, 
169 (1998). 


\bibitem{frad}
E. Fradkin, {\it Field Theories of Condensed Matter Systems}, 
Addison-Wesley, Redwood City, (1991).

\bibitem{size} 
Modestly large {\it does not\/} imply the classical limit, i.e., 
values of $S=2$ or $3$ will exhibit the effect albeit with a limit 
region over which the physical properties are quasi-periodic in the 
magnetization.

\bibitem{chud} For a recent review of the model and theory, see: E. M. 
Chudnovsky in {\it Quantum tunneling of Magnetization - QTM '94}, pg.  
1-18, eds.  L.~Gunther and B.~Barbara, Kluwer Academic Publishers, 
Dordrech 1995.

\bibitem{seb} See, e.g., S.~E. Barnes, Advances in Physics {\bf 30} 
801-938 (1981).

\bibitem{garg}
A. Garg, Europhys. Lett., {\bf 22}, 205 (1993).



\bibitem{sebetal}  S. E. Barnes, B. Barbara, R. Ballou and J. Strelen, Phys.
Rev. Lett. {\bf 79}, 289 (1997).


\bibitem{gap} This fixed point corresponds to a rather trivial case of 
a ``Haldane'' gap for integer spin since this is a single spin effect 
which arises very directly from the anisotropy term.  In fact it is 
known that this ``large-D'' phase is separated from the zero field 
``Haldane'' phase by a critical value $D_{c} = 2J$, see e.g., ref.  
5 for discussion and references.  However, as shown in Refs.  2 and 
3, precisely for the case of a single spin, the presence of such a gap 
for an integer single spin and its absence for half-integer spin is of 
the same topological origin as is the Haldane conjecture, i.e., the 
present fixed point for zero field {\it does\/} have a Haldane gap 
albeit of a rather trivial origin.

\bibitem{Tsuneto} T. Tsuneto and T. Murao, Physica {\bf 51}, 186 (1971).

\bibitem{Tachiki} M. Tachiki, T. Yamada, and S. Maekawa, {\bf 29}, 656 
(1970).

\bibitem{bog} N. M. Bogoliubov, A. G. Izergin and V. E. Korepin, 
Nuclear Phys. B {\bf 275} 687 (1986).

\bibitem{taft} K. L. Taft, C. D. Delfs, G. C. Papaefthymiou, S. Foner, 
D. Gatteschi and S. J. Lippard, J. Amer. Chem. Soc. {\bf 116}, 823 
(1994).

\bibitem{thomas} L. Thomas, F. Lionti, R. Ballou, R. Sessoli, A. 
Caneschi and B. Barbara, Nature {\bf 383}, 145 (1996); J. R. Friedman 
et al., Phys.  Rev.  Lett.  {\bf 76}, 3820 (1996).

\end{references}
\end{document}